\documentclass[12pt, a4paper]{article}
\usepackage{cite}
\usepackage{amsmath,amssymb}
\input{colordvi.tex}
\usepackage{color}
\usepackage{graphicx}
\usepackage{comment}
\usepackage[colorlinks=true,linkcolor=red,urlcolor=blue,citecolor=blue]{hyperref}

\begin{document}

%%%%%%%%%%%%%%%%%%%%%%%%%%%%%%%%%%%%%%%%%%%%%%%%%%
\begin{titlepage}

\begin{center}

\hfill UT-18-08\\
\hfill KEK-TH-2047\\

\vskip .75in

{\Large \bf 
Production of Purely Gravitational Dark Matter
}

\vskip .75in

{\large
Yohei Ema$^{(a,b)}$, Kazunori Nakayama$^{(a,c)}$ and Yong Tang$^{(a)}$
}

\vskip 0.25in

$^{(a)}${\em Department of Physics, Faculty of Science,\\
The University of Tokyo,  Bunkyo-ku, Tokyo 113-0033, Japan}\\[.3em]
$^{(b)}${\em Theory Center, High Energy Accelerator Research Organization (KEK),\\
Tsukuba, Ibaraki 305-0801, Japan}\\[.3em]
$^{(c)}${\em Kavli IPMU (WPI), UTIAS,\\
The University of Tokyo,  Kashiwa, Chiba 277-8583, Japan}

\end{center}
\vskip .5in

\begin{abstract}

In the purely gravitational dark matter scenario, the dark matter particle does not have any interaction except for gravitational one.
We study the gravitational particle production of dark matter particle in such a minimal setup and show that correct amount of dark matter can be produced depending on the inflation model and the dark matter mass. In particular, we carefully evaluate the particle production rate from the transition epoch to the inflaton oscillation epoch in a realistic inflation model
and point out that the gravitational particle production is efficient even if dark matter mass is much larger than the Hubble scale
during inflation as long as it is smaller than the inflaton mass.

\end{abstract}

\end{titlepage}

%\tableofcontents

\renewcommand{\thepage}{\arabic{page}}
\setcounter{page}{1}
\renewcommand{\thefootnote}{\#\arabic{footnote}}
\setcounter{footnote}{0}
%%%%%%%%%%%%%%%%%%%%%%%%%%%%%%%%%%%%%%%%%%%%%%%%%%

\newpage

%%%%%%%%%%%%%%%%%%%%%%%%%%%%%%%%%%%%%%%%%%%%%%%%%%
\section{Introduction}
\label{sec:Intro}
%%%%%%%%%%%%%%%%%%%%%%%%%%%%%%%%%%%%%%%%%%%%%%%%%%

There are many particle physics models to explain cosmological dark matter (DM).
Most models introduce (small) DM interactions with some other fields to obtain the desired DM abundance.
In this respect, the simplest possibility is that DM interacts with other sectors only through gravity.
We call it as the purely gravitational DM (PGDM). Interestingly, even in the case where DM interacts only through gravity, there are several processes which contribute to the DM particle production.

The so-called gravitational particle production often refers to the particle creation due to the 
expansion of the Universe, 
or the time dependence of the cosmic scale factor~\cite{Parker:1969au,Birrell:1982ix}.
The classic paper~\cite{Ford:1986sy} studied the scalar particle production during the transition from inflation to the matter dominated (MD) or radiation dominated (RD) Universe and found that number density of $n_\chi \sim H_{\rm inf}^3$ is produced, where $H_{\rm inf}$ denotes the Hubble scale during inflation, if the scalar particle $\chi$ is effectively massless or $m_\chi \ll H_{\rm inf}$ with $m_\chi$ being the mass of $\chi$.
Refs.~\cite{Chung:1998zb, Chung:2001cb} considered the gravitational particle creation at the end of inflation as a mechanism for the production of supermassive DM. More recently, Refs.~\cite{Ema:2015dka, Ema:2016hlw} pointed out that gravitational particle production always happens during the inflaton oscillation era after inflation, since the cosmic scale factor is also oscillating. There it is found that the production is most efficient at early epoch and the produced number density is $n_\chi \sim H_{\rm inf}^3$, similar to that found in Ref.~\cite{Ford:1986sy}.
It becomes clear that this estimation applies to $m_\chi$ as large as the inflaton mass $m_\phi$, even if it is much larger than $H_{\rm inf}$.
It may be easily understood once one regards the gravitational particle production due to the oscillating background as the ``gravitational annihilation'' of inflaton.\footnote{
	Refs.~\cite{Bassett:1997az,Tsujikawa:1999jh} pointed out that there is a tachyonically resonant production
	if $\chi$ has a (large) non-minimal coupling to gravity. 
	See also Refs.~\cite{Markkanen:2015xuw}.
}

On the other hand, Refs.~\cite{Garny:2015sjg,Tang:2016vch,Tang:2017hvq,Garny:2017kha} studied the production of PGDM
through the annihilation of standard model particles in thermal bath with s-channel graviton exchange. It may also be regarded as a kind of gravitational particle creation. The efficiency of DM production is dominated at earlier epoch or high cosmic temperature.
However, in the very early epoch the Universe is dominated by the (coherently oscillating) inflaton field, hence the DM production is expected to be dominated by the annihilation of inflaton rather than particles in thermal bath in most cases~\cite{Tang:2017hvq}.

In this paper we revisit the gravitational particle production in a realistic situation.  It is often misunderstood that the gravitational particle production is not efficient if $m_\chi \gg H_{\rm inf}$. It is not always true, however. In many inflation models, there is a large hierarchy between the inflaton mass scale and Hubble scale: $m_\phi \gg H_{\rm inf}$. Gravitational particle production is efficient even for $m_\phi > m_\chi \gg H_{\rm inf}$. This point is not stressed in literature except for a few works~\cite{Ema:2015dka, Ema:2016hlw}.
Therefore we want to study the gravitational production in a comprehensive manner.

This paper is organized as follows. In Sec.~\ref{sec:scalar} some basics of a scalar field in the expanding background are described
and equations to estimate the gravitational particle production are derived. In Sec.~\ref{sec:prod} we (semi)-analytically and numerically evaluate the gravitational particle production rate in a realistic inflationary cosmology. Sec.~\ref{sec:conc} is devoted to summary and discussion.

%%%%%%%%%%%%%%%%%%%%%%%%%%%%%%%%%%%%%%%%%%%%%%%%%%
\section{Scalar field in cosmological background}
\label{sec:scalar}
%%%%%%%%%%%%%%%%%%%%%%%%%%%%%%%%%%%%%%%%%%%%%%%%%%

%%%%%%%%%%%%%%%%%%%%%%%%%%%%%%%%
\subsection{Model and equations of motion}
%%%%%%%%%%%%%%%%%%%%%%%%%%%%%%%%

Let us consider an action
\begin{align}
	S = \int d^4 x \sqrt{-g} \left( \frac{1}{2}(M_P^2 - \xi \chi^2)R - \frac{1}{2}g^{\mu\nu}\partial_\mu\phi \partial_\nu\phi - V(\phi)
	- \frac{1}{2}g^{\mu\nu}\partial_\mu\chi \partial_\nu\chi - \frac{1}{2} m_\chi^2 \chi^2 \right),
\end{align}
where $M_P$ is the reduced Planck scale, $R$ is the Ricci scalar,
$\phi$ denotes the inflaton field with $V(\phi)$ being its potential and $\chi$ denotes a real scalar field.
It has a $Z_2$ symmetry under which $\chi$ changes its sign, and hence $\chi$ is stable and is a candidate of DM.
We assume that $\chi$ does not have a direct coupling to the inflaton and other standard model fields.
It interacts only through the metric or the gravity.
The coupling strength to the gravity is controlled by the non-minimal coupling $\xi$.
Pure Einstein gravity corresponds to $\xi=0$ and the conformal coupling corresponds to $\xi=1/6$.

We use the Friedmann-Robertson-Walker metric:
\begin{align}
	g_{\mu\nu} dx^\mu dx^\nu = a^2(\tau)(-d\tau^2 + d\vec x^2),
\end{align}
where $a(\tau)$ denotes the cosmic scale factor with $\tau$ being the conformal time, which is related to the physical time as $dt = a d\tau$.
Defining $\widetilde\chi \equiv a\chi$, the action of $\widetilde\chi$ is given by
\begin{align}
	S = \int d\tau d^3x \frac{1}{2}\left[ \widetilde\chi'^2 - (\partial_i\widetilde\chi)^2 - m^{\rm (eff)2}_\chi \widetilde\chi^2 \right],
	~~~~~~m^{\rm (eff)2}_\chi \equiv a^2m_\chi^2-(1-6\xi)\frac{a''}{a},
	\label{S_chi}
\end{align}
where the prime denotes the derivative with respect to $\tau$.
Thus $\widetilde\chi$ satisfies the equation of motion
\begin{align}
	\widetilde\chi'' - \partial^2_i \widetilde \chi + m^{\rm (eff)2}_\chi \widetilde\chi = 0.
\end{align}

Treated as classical background,  $\phi$ has the following equation of motion,
\begin{align}
	\ddot\phi + 3H\dot \phi + \frac{\partial V}{\partial\phi}=0,   \label{eom_phi}
\end{align}
where the dot denotes the derivative with respect to the physical time $t$ and the Hubble parameter $H$ is given by the Friedmann equation,
\begin{align}
	H^2 = \left( \frac{\dot a}{a}\right)^2 = \frac{1}{3M_P^2}\left(\frac{1}{2}\dot\phi^2 + V(\phi) \right),   \label{Fried}
\end{align}
with the $\chi$ contribution to the energy density neglected. Thus for any given inflation model, we can calculate the production rate of $\chi$ through the time dependence of the scale factor $a$ in (\ref{S_chi}). These equations are written in terms of the conformal time as
\begin{align}
	\phi'' + 2\mathcal H \phi' + a^2\frac{\partial V}{\partial \phi} = 0,  \label{phi_eom}
\end{align}
where the conformal Hubble parameter $\mathcal H$ is
\begin{align}
	\mathcal H^2 \equiv \left(\frac{a'}{a}\right)^2
	= \frac{1}{3M_P^2}\left(\frac{1}{2}\phi'^2 + a^2 V \right).  \label{H2}
\end{align}
The Friedmann equation of the second kind is given by
\begin{align}
	\frac{a''}{a} 
	= \frac{a^2 R}{6}
	= \frac{4a^2 V - \phi'^2}{6M_P^2}.
	\label{Ricci}
\end{align}
%%

%%%%%%%%%%%%%%%%%%%%%%%%%%%%%%%%
\subsection{Quantization and adiabatic vacuum}
%%%%%%%%%%%%%%%%%%%%%%%%%%%%%%%%

Since $m^{\rm (eff)2}_\chi$ is time-dependent in the expanding Universe, we should be careful about the choice of 
mode function and vacuum state. Let us define the Fourier mode as
\begin{align}
	\widetilde\chi(\tau,\vec x) = \int \frac{d^3 k}{(2\pi)^3} \left[a_{\vec k} \chi_ k(\tau) + a^\dagger_{-\vec k} \chi_k^*(\tau) \right] e^{i \vec k\cdot \vec x},
	\label{Fourier}
\end{align}
where $\chi_k =\chi_{-k}$ must be satisfied from the reality of $\chi$, 
and the ladder operator satisfies
\begin{align}
	\left[a_{\vec k},a_{\vec k'}^\dagger \right] = (2\pi)^3\delta(\vec k- \vec k'),~~~~~
	\left[a_{\vec k},a_{\vec k'} \right]= \left[a_{\vec k}^\dagger,a_{\vec k'}^\dagger \right]=0.
\end{align}
The Fourier mode satisfies the equation of motion:
\begin{align}
	\chi_k'' + \omega_k^2 \chi_k = 0,~~~~~\omega_k^2 \equiv k^2 + m^{\rm (eff)2}_\chi.   \label{eom_chi}
\end{align}
From the canonical commutation relation
\begin{align}
	\left[ \widetilde\chi(\vec x), \widetilde\chi'(\vec x') \right] = i \delta(\vec x - \vec x'),
\end{align}
we obtain the normalization condition
\begin{align}
	\chi_k \chi_k'^* - \chi_k^*\chi_k' = i.   \label{chikchik}
\end{align}
The vacuum state $\left |0\right>$ is defined as $a_{\vec k} \left|0\right>=0$ 
for some mode function $\chi_k$ at some initial time $\tau=\tau_i$. 
In the Heisenberg picture, the state does not evolve once we fix it at the initial time.
Instead, the mode function develops with time, which may be interpreted as the particle production as will be shown later.

Given initial conditions, one can directly solve (\ref{eom_chi}), but here we use a different technique~\cite{Kofman:1997yn}.
Let us rewrite $\chi_k(\tau)$ as follows:
\begin{align}
	\chi_k(\tau) = \alpha_k(\tau) v_k(\tau) + \beta_k(\tau) v_k^*(\tau),~~~~~~v_k(\tau) \equiv \frac{1}{\sqrt{2\omega_k}}\exp\left(-i\int \omega_k d\tau \right).
\end{align}
Since both $\alpha_k(\tau)$ and $\beta_k(\tau) $ are time-dependent, one can always write $\chi_k$ in this form.
Still there is a degree of freedom for the choice of $\alpha_k(\tau)$ and $\beta_k(\tau)$.
One can impose the following condition, which is consistent with the equation of motion~(\ref{eom_chi}):
\begin{align}
	\alpha_k' v_k  = \frac{\omega_k'}{2\omega_k}v_k^* \beta_k,~~~~~~
	\beta_k' v_k^* = \frac{\omega_k'}{2\omega_k}v_k \alpha_k.   \label{alpha}
\end{align}
Instead of solving (\ref{eom_chi}), one may solve (\ref{alpha}), which is often much easier for the purpose of evaluating the particle production numerically.
Note that (\ref{chikchik}) ensures 
\begin{align}
	\left|\alpha_k(\tau)\right|^2 - \left|\beta_k(\tau)\right|^2 = 1.
\end{align}

In order to extract the number density from $\alpha_k$ and $\beta_k$,
we must first specify the initial condition for the mode function 
(which corresponds to the choice of the vacuum state) 
and the observer state which counts the number density.
In general, there is no preferred choice for the states of the vacuum and the observer
in curved spacetime.
In our case, however, we may formally assume that the spacetime is asymptotically static
in the far past $\tau \rightarrow -\infty$ (deep in the inflationary era) 
as well as in the far future $\tau \rightarrow +\infty$ (deep in the MD or RD era).
In such a case, it is natural to take the vacuum/observer as
the negative frequency modes in the limit $\tau\rightarrow -\infty/\infty$, respectively.
The negative frequency mode approaches to
\begin{align}
	\chi_k(\tau) \rightarrow \frac{1}{\sqrt {2k}} e^{-ik\tau},
	\label{eq:init_cond}
\end{align}
in the limit $\tau \rightarrow -\infty$.
Thus we take $\alpha(\tau_i) = 1$ and $\beta(\tau_i) = 0$ in the limit $\tau_i \rightarrow -\infty$
as the initial condition.
Note that it represents the adiabatic vacuum of the infinite order~\cite{Birrell:1982ix},
since the spacetime is assumed to be static in the far past/future regions.\footnote{
	One can show that our definition of $f_\chi$ is equivalent to 
	the number density defined by the Bogoliubov coefficient
	where the vacuum and the observer states are taken as 
	the zero-th order adiabatic vacuum
	as long as $\omega_k'/\omega_k = 0$ at $\tau = \tau_i$ and $\tau_f$,
	where $\tau_f$ is the conformal time at which the number density is evaluated.
	Since the adiabatic expansion is exact in the limit in the far past/future regions,
	our $f_\chi$ coincides with the number density 
	defined by the adiabatic vacuum of the infinite order
	for $-\tau_i, \tau_f \rightarrow \infty$ as well.
} In our numerical calculation, however, it is of course impossible to run from $\tau_i = -\infty$,
and hence we start our numerical calculation 
with $\alpha(\tau_i) = 1$ and $\beta(\tau_i) = 0$ at some large but finite $\tau_i$.
We will discuss how to infer the result with $\tau_i \rightarrow -\infty$
from our numerical results with finite $\tau_i$ in the next section.

Here is a comment on the size of $m_\chi^2$ and $\xi$.
The exact solution to (\ref{eom_chi}) during the (pure) de Sitter era
with the initial condition~\eqref{eq:init_cond} is given by
\begin{align}
	\chi_k(\tau) = e^{\frac{i(2\nu+1)\pi}{4}}\frac{1}{\sqrt{2k}}\sqrt{\frac{-\pi k\tau}{2}} H_{\nu}^{(1)}(-k\tau),~~~~~~
	\nu^2 \equiv \frac{9}{4} - 12\xi - \frac{m_\chi^2}{H^2},
	\label{chik_exact}
\end{align}
where $H_{\nu}^{(1)}$ denotes the Hankel function of the first kind.
In this paper we mainly concentrate on the case of $\nu^2<1/4$.
Otherwise long wave superhorizon fluctuation develops and $\chi$ obtains a homogeneous classical field value,\footnote{
	The long wavelength limit is $H_{\nu}^{(1)}(-k\tau) \to -(i/\pi) \Gamma(\nu) (-k\tau/2)^{-\nu}$ for $k\tau \to 0$.
	Note that $m_{\chi}^{\rm (eff)2} > 0$ always holds if $\nu^2  <1/4$ 
	during inflation and hence $\omega_k$ is real for all the wavenumber $k$.
}
and we must take account of the coherent oscillation of the homogeneous $\chi$ field for estimation of the final DM abundance.
For the small nonminimal coupling $|\xi| \leq 1/6$, $\nu^2<1/4$ roughly means that $\chi$ is heavier than the Hubble scale during and after inflation.
For the nonminimal coupling $\xi \geq 1/6$, $\nu^2<1/4$ always holds even if $m_\chi^2 \ll H^2$ as far as $m_\chi^2$ is positive.

%%%%%%%%%%%%%%%%%%%%%%%%%%%%%%%%
\subsection{Energy and number density}
%%%%%%%%%%%%%%%%%%%%%%%%%%%%%%%%

In order to evaluate the amount of created particles in the cosmological background, one must define the energy or number density.
Of course, they are actually UV divergent and hence we need to carefully renormalize them.
The energy momentum tensor of $\chi$ is given by $\sqrt{-g}T^{\chi}_{\mu\nu}=-2\delta (\sqrt{-g}\mathcal L_\chi) /\delta g_{\mu\nu}$~\cite{Birrell:1982ix}:
\begin{align}
	T_{\mu\nu}^\chi =& (1-2\xi) \partial_\mu\chi \partial_\nu\chi + \left(2\xi-\frac{1}{2}\right)g_{\mu\nu}(\partial\chi)^2
	+2\xi\left( g_{\mu\nu} \chi \Box \chi - \chi \nabla_\mu \partial_\nu \chi\right)  \nonumber\\
	& + \xi G_{\mu\nu} \chi^2 - \frac{1}{2} m_\chi^2 g_{\mu\nu} \chi^2,
\end{align}
where $G_{\mu\nu}$ denotes the Einstein tensor and partial derivative is taken with respect to physical time $t$. 
The energy density may be defined through $\rho_\chi = T_{00}^\chi$. Using (\ref{Fourier}), we obtain
\begin{align}
	a^4 \rho_\chi &= \int \frac{d^3k}{(2\pi)^3}\frac{1}{2}\left[ |\chi_k'|^2 +(k^2 + a^2 m_\chi^2)|\chi_k|^2 
	+\left(1-6\xi\right)\left\{\mathcal H^2 |\chi_k|^2-\mathcal H(\chi_k {\chi_k^*}' +\chi_k^*\chi_k'  )  \right\} \right]\\
	&= \int \frac{d^3k}{(2\pi)^3}\frac{\omega_k}{2}\left[  \left(1+2|\beta_k|^2\right) \left(1+ (1-6\xi)\frac{2\mathcal H^2 + \mathcal H'}{2\omega_k^2}\right)
	+(1-6\xi)\left( \frac{1-2i \mathcal H\omega_k}{\omega_k}
	\alpha_k \beta_k^* v_k^2 + {\rm h.c.}  \right)	\right].
\end{align}
If we naively introduce a physical UV cutoff scale $\Lambda$, we find terms like $\rho_\chi \propto \Lambda^4$, $H^2 \Lambda^2$
and $H^4 \log \Lambda$.
These divergences are renormalized by the constant term, Planck constant and the coefficient of the $R^2$ term in the action, 
respectively~\cite{Birrell:1982ix}.\footnote{
	As for the quadratic divergence, one can explicitly check that a term like $\dot H \Lambda^2$ is cancelled out.
}
Since we want to calculate the energy density of $\chi$ in the far future in which $\mathcal H\to 0$ and $\mathcal H' \to 0$,
one can safely define the renormalized energy density as 
\begin{align}
	a^4(\tau) \rho_\chi^{\rm (ren)}(\tau)&= \int \frac{d^3k}{(2\pi)^3}\omega_k f_\chi(k,\tau),   ~~~~~~ f_\chi(k,\tau) \equiv |\beta(k,\tau)|^2,\label{fchi}
\end{align}
where $f_\chi(k)$ denotes the phase space density of the $\chi$ particle with physical momentum $k/a$. Similarly, the number density is given by
\begin{align}
	a^3(\tau) n_\chi(\tau) = \int \frac{d^3k}{(2\pi)^3}f_\chi(k,\tau).
\end{align}
Therefore, given an cosmological model, we can calculate the evolution $\beta_k$ by using (\ref{alpha})
and then estimate the number of produced particles. In particular, in the most cases of our interest $\beta_k$ is so small that we can safely take $\alpha_k\simeq 1$.
Then we have
\begin{align}
	\beta_k(\tau_f) \simeq \int_{\tau_i}^{\tau_f} \frac{\omega_k'}{2\omega_k}e^{-2i\Omega_k(\tau)}d\tau,
	~~~~~~\Omega_k(\tau) \equiv \int_{\tau_i}^\tau\omega_k d\tau.
	\label{betak}
\end{align}
As we discussed in the previous subsection, we take the limit $\tau_i \rightarrow -\infty$,
and evaluate $f_\chi$ at $\tau_f \rightarrow \infty$ where (\ref{fchi}) is rigorous,
assuming that the spacetime is static in these regions.

%%%%%%%%%%%%%%%%%%%%%%%%%%%%%%%%%%%%%%%%%%%%%%%%%%
\section{Production of purely gravitational dark matter}
\label{sec:prod}
%%%%%%%%%%%%%%%%%%%%%%%%%%%%%%%%%%%%%%%%%%%%%%%%%%

%%%%%%%%%%%%%%%%%%%%%%%%%%%%%%%%
\subsection{Analytic estimation}
%%%%%%%%%%%%%%%%%%%%%%%%%%%%%%%%

%%%%%%%%%%%%%%%%%%%%%%%%%%%%%%%%
\subsubsection{Conformal coupling} \label{sec:conf}
%%%%%%%%%%%%%%%%%%%%%%%%%%%%%%%%

Now let us evaluate $\beta_k$ through (\ref{betak}).
In the following we always assume $m_\chi \lesssim m_\phi$.
First, let us consider the case of conformal coupling $\xi=1/6$ for simplicity.
Of course, in the massless limit $m_\chi=0$ the $\chi$ field does not feel the Universe expanding due to the conformality
and no particle production occurs. However, for finite $m_\chi$, particle production happens.
We can approximate $\Omega_k$ as
\begin{align}
	\Omega_k(\tau) \simeq 
	\begin{cases}
		k\tau & {\rm for}~~~k^2 \gg a^2 m_\chi^2 \\
		m_\chi t & {\rm for}~~~k^2 \ll a^2 m_\chi^2
	\end{cases},
\end{align}
where we have ignored additional irrelevant constant. The integrand is given by
\begin{align}
	\frac{\omega_k'}{2\omega_k}  = \frac{\mathcal H}{2}\frac{a^2 m_\chi^2}{k^2 + a^2 m_\chi^2}
	\simeq \begin{cases}
		\displaystyle\frac{\mathcal Ha^2 m_\chi^2}{2k^2} & {\rm for}~~~k^2 \gg a^2 m_\chi^2 \\
		\displaystyle\frac{\mathcal H}{2}                           & {\rm for}~~~k^2 \ll a^2 m_\chi^2
	\end{cases}.
\end{align}
The integrand approaches to zero both in the limit $\tau\to-\infty$ (deep in the inflationary era) 
and $\tau\to + \infty$ (deep in the MD or RD era).
Thus the integration (\ref{betak}) is well-defined,
\begin{align}
	\beta_k(\tau_f) \simeq \int_{\tau_i}^{\tau_k} \frac{\mathcal Ha^2 m_\chi^2}{2k^2}e^{-2ik\tau} d\tau +
	\int_{t_k}^{t_f} \displaystyle\frac{H}{2}e^{-2im_\chi t}dt,
	\label{betak_conf}
\end{align}
where $\tau_k$ is defined as the conformal time at which $a(\tau_k) m_\chi = k$.
Now we need the time dependence of the scale factor $a(t)$ for evaluating this integral.
As explicitly noticed in Ref.~\cite{Ema:2015dka}, the scale factor $a(t)$ contains rapidly oscillating part during the inflaton coherent oscillation,
as 
\begin{align}
	a(t) \simeq \left< a(t)\right>\left(1 - \frac{(\phi-v_\phi)^2}{8M_P^2}\right),   \label{a_osc}
\end{align}
where $v_\phi$ is the final VEV of inflaton and $\left< a(t)\right>$ is the time-averaged value over the inflaton oscillation period.
The particle production may be caused by both the time-dependence of the averaged value $\left<a(t)\right>$
and also by the rapidly oscillating part.
We stress that such a clear separation is possible only in the harmonic inflaton oscillation regime and distinction between
``slowly changing'' part and ``rapidly changing'' part is not well defined in general.
Nevertheless, this picture helps the analytic understandings of the particle production.

First let us focus on the effect by the time-averaged part $\left<a(t)\right>$.
We discuss this contribution in detail in App.~\ref{sec:toy} and the resultant phase space distribution is given by~(\ref{fk_toy_lowk}) and~(\ref{fk_toy_highk}):
\begin{align}
	f_\chi(k) \sim \begin{cases}
	\displaystyle\exp\left( -\frac{cm_\chi}{H_{\rm inf}} \right) & {\rm for}~~~k < a_{\rm end} m_\chi  \\
	\displaystyle\frac{a_{\rm end}^4 m_\chi^4}{k^4}\exp\left( -\frac{ck}{a_{\rm end}H_{\rm inf}} \right)
	+ \exp\left[ -c\left(\frac{k}{k_c} \right)^{3/2} \right]  & {\rm for}~~~k > a_{\rm end} m_\chi
	\end{cases},
	\label{f_conf_adi}
\end{align}
where $c\sim 5$ and $a_{\rm end}$ denotes the scale factor at the end of inflation and
$k_c$ is defined by the momentum so that $m_\chi t_k = 1$.

Next we consider the effect from the rapidly oscillating part in (\ref{a_osc}).
The particle production due to such a rapidly oscillating effective mass term is analogous to the process known as the narrow resonance due to the oscillating scalar field~\cite{Dolgov:1989us,Traschen:1990sw,Shtanov:1994ce,Kofman:1997yn}.
In our situation the parametric resonance effect is neglected since the Hubble expansion is fast enough to  
expel the created particles from the resonance band in the momentum space~\cite{Ema:2015dka}
and in such a case the production rate is the same as the perturbative decay/annihilation rate.
The effective inflaton annihilation rate in this case is
\begin{align}
	\Gamma_{\phi\phi\to\chi\chi} \simeq \frac{C}{16\pi} \frac{\Phi^2}{M_P^4}\frac{m_\chi^4}{m_\phi},  \label{Gamma_conf}
\end{align}
if $m_\chi \lesssim m_\phi$, where $\Phi$ denotes the oscillation amplitude of the inflaton field and $C$ is a numerical factor.
Thus the created number density during one Hubble time is given by
\begin{align}
	n_\chi \simeq \frac{9C}{4\pi} \left( \frac{m_\chi}{m_\phi} \right)^4 H^3.
\end{align}
One immediately notices that the dominant contribution comes from the earliest epoch, i.e., $H\sim H_{\rm inf}$.
The typical physical momentum of $\chi$ is $k/a \sim m_\phi$ for each epoch.
The final momentum distribution of $\chi$ is not exponentially suppressed for large $k$ since 
the production continues until the inflaton decays.
We obtain the following approximated form:
\begin{align}
	f_\chi(k,\tau) \sim \begin{cases}
		\displaystyle \left(\frac{H_{\rm inf}}{m_\phi} \right)^3 \left( \frac{m_\chi}{m_\phi} \right)^4&~~~{\rm for}~~~ k < a_{\rm end} m_\phi, \\
		\displaystyle \left(\frac{H_{\rm inf}}{m_\phi} \right)^3 \left( \frac{m_\chi}{m_\phi} \right)^4\left(\frac{k}{ a_{\rm end}m_\phi} \right)^{-9/2} 
		&~~~{\rm for}~~~ k > a_{\rm end} m_\phi,
	\end{cases} 
	\label{f_conf_osc}
\end{align}
and there will be an exponential cutoff at $k \sim a(t=\Gamma_{\rm inf}^{-1}) m_\phi$
where $\Gamma_{\rm inf}$ denotes the total decay width of the inflaton.\footnote{
	For the completion of the reheating, we need to introduce inflaton coupling to some other sector.
}
This contribution is not suppressed even for $m_\chi \gg H_{\rm inf}$.
Note also that the low momentum behavior of (\ref{f_conf_osc}) may not be so simple because of the nontrivial time scale of the inflaton dynamics
during the transition from inflation to the reheating era.
The final momentum distribution is roughly the sum of (\ref{f_conf_osc}) and (\ref{f_conf_adi}).
In realistic situation, as we will see later, these two contribution should be smoothly connected 
because the ``end'' of inflation is rather vague and the inflaton oscillation is far from harmonic at the early stage of the reheating
and its typical time scale changes from $H_{\rm inf}$ to $m_\phi$ gradually. 
The number density is then given by
\begin{align}
	n_\chi(\tau) \sim H_{\rm inf}^3 \left( \frac{a_{\rm end}}{a(\tau)} \right)^3 \left[
		\mathcal C \left( \frac{m_\chi}{m_\phi} \right)^4 + \eta \frac{m_\chi}{H_{\rm inf}}
	\right],
	\label{nchi_conf}
\end{align}
where we will numerically find $\mathcal C \sim 10^{-3} - 10^{-2}$ and
\begin{align}
	\eta \equiv \mathcal A \times \begin{cases}
	\displaystyle 1  &~~~{\rm for}~~~m_\chi < H_{\rm inf}, \\
	\displaystyle \frac{m_\chi^2}{H_{\rm inf}^2} e^{-cm_\chi / H_{\rm inf}} 
		&~~~{\rm for}~~~ m_\chi >H_{\rm inf}.
	\end{cases}
\end{align}
with $\mathcal A \sim 6\times 10^{-4}$ for $m_\chi \lesssim H_{\rm inf}$.
Thus the contribution from the pure Hubble expansion likely to dominate for $m_\chi < H_{\rm inf}$
while that from the oscillation effect (\ref{f_conf_osc}) becomes important for $m_\chi > H_{\rm inf}$.
In the latter case, we can evaluate the present DM energy density from the gravitational production divided by the entropy density as
\begin{align}
	\frac{\rho_\chi^{\rm (GP)}}{s} 
	&\sim \frac{\mathcal C}{4} \frac{m_\chi H_{\rm inf}T_{\rm R}}{M_P^2}\left( \frac{m_\chi}{m_\phi} \right)^4 \nonumber\\
	&\simeq 4\times 10^{-10}\,{\rm GeV}\,\mathcal C\left(\frac{m_\chi}{10^9\,{\rm GeV}}\right)
	\left(\frac{H_{\rm inf}}{10^9\,{\rm GeV}}\right)
	\left(\frac{T_{\rm R}}{10^{10}\,{\rm GeV}}\right)\left( \frac{m_\chi}{m_\phi} \right)^4,
	\label{rho_GP_conf}
\end{align}
for $m_\chi < m_\phi$, where $T_{\rm R}$ denotes the reheating temperature after inflation
and we assumed that the inflaton coherent oscillation behaves as non-relativistic matter.

%%%%%%%%%%%%%%%%%%%%%%%%%%%%%%%%
\subsubsection{Minimal coupling}   \label{sec:minimal}
%%%%%%%%%%%%%%%%%%%%%%%%%%%%%%%%

Now let us consider the case of minimal coupling, $\xi=0$.
In this case, we have
\begin{align}
	m_{\chi}^{\rm (eff)2} = a^2 \left( m_\chi^2 -\frac{R}{6}\right),
\end{align}
and we obtain
\begin{align}
	\frac{\omega_k'}{2\omega_k} \simeq \frac{1}{2}\frac{\mathcal H a^2 m_\chi^2 - (a^2 R)' /12}{k^2 + a^2 m_\chi^2}.
\end{align}
where we have assumed $m_\chi \gtrsim H_{\rm inf}$ to avoid the growth superhorizon modes as already mentioned.
The first term is the same as the conformal coupling case and hence we focus on the second term hereafter.
In the following we also assume $m_\chi \lesssim m_\phi$.

The Ricci scalar $R$ is expressed in terms of $\phi$ as  (\ref{Ricci}) and hence it is a rapidly oscillating function when the inflaton oscillates coherently.
Therefore we can divide $R$ into its oscillation averaged part and the oscillating part, schematically as $R \sim \left<R\right> + m_\phi^2(\phi-v_\phi)^2/M_P^2$.
First let us consider the effect due to the averaged part. After the partial integration, the integral (\ref{betak}) is expressed as
\begin{align}
	\beta_k \sim \frac{-i}{12}\int \frac{a^2 \left<R\right>}{\omega_k}e^{-2i\Omega_k} d\tau.
\end{align}
It may be dominated around the transition $\tau\sim \tau_{\rm end}$ where $a^2 R / \omega_k$ takes its maximum value.
Thus we roughly have\footnote{
	Even for $m_\chi \ll H_{\rm inf}$, the result for the high momentum mode $k > a_{\rm end }H_{\rm inf}$ looks similar
	to that for the high momentum limit of the case of $m_\chi \gtrsim H_{\rm inf}$.
	These high momentum contribution goes as $n_\chi \sim H_{\rm inf}^3 (a_{\rm end} / a(t))^3$, as given in Ref.~\cite{Ford:1986sy},
	but the long wavelength contribution can be dominant in such a case. See Sec.~\ref{sec:conc}.  
}
\begin{align}
	f_\chi(k) \sim \begin{cases}
	\displaystyle  \frac{H_{\rm inf}^2}{m_\chi^2}\exp\left(-\frac{cm_\chi}{H_{\rm inf}}\right)
	& {\rm for}~~~k < a_{\rm end} m_\chi  \\
	\displaystyle  \frac{a_{\rm end}^2 H_{\rm inf}^2}{k^2}\exp\left(-\frac{ck}{a_{\rm end }H_{\rm inf}}\right)  
	& {\rm for}~~~k > a_{\rm end} m_\chi
	\end{cases}.
	\label{f_min_adi}
\end{align}

Next we consider a particle production due to the oscillating part in $R$.
As already mentioned in the case of conformal coupling, such a rapidly oscillating term induces particle production similar to the standard narrow resonance.
On these grounds, the effective inflaton annihilation rate during the inflaton coherent oscillation is given by
\begin{align}
	\Gamma_{\phi\phi\to\chi\chi} \simeq \frac{C}{16\pi} \frac{\Phi^2}{M_P^4}m_\phi^3,
\end{align}
if $m_\chi \lesssim m_\phi$ where $C$ is a numerical factor and $\Phi$ denotes the inflaton oscillation amplitude.
Thus the created number density during one Hubble time is given by
\begin{align}
	n_\chi \simeq \frac{9C}{4\pi} H^3.
\end{align}
Again, the dominant contribution comes from the earliest epoch, i.e., $H\sim H_{\rm inf}$.
Similarly to the case studied in the conformal coupling case, the typical physical momentum of $\chi$ is $k/a \sim m_\phi$ for each epoch
and the final momentum distribution of $\chi$ is not exponentially suppressed for large $k$ since 
the production continues until the inflaton decays.
We expect that the momentum distribution looks like
\begin{align}
	f_\chi(k,\tau) \sim \begin{cases}
		\displaystyle \left(\frac{H_{\rm inf}}{m_\phi} \right)^3 &~~~{\rm for}~~~ k < a_{\rm end} m_\phi, \\
		\displaystyle \left(\frac{H_{\rm inf}}{m_\phi} \right)^3 \left(\frac{k}{ a_{\rm end}m_\phi} \right)^{-9/2} 
		&~~~{\rm for}~~~ k > a_{\rm end} m_\phi,
	\end{cases} 
	\label{f_minimal}
\end{align}
and there is an exponential cutoff at $k \sim a(t=\Gamma_{\rm inf}^{-1}) m_\phi$. It is not suppressed even for $m_\chi \gg H_{\rm inf}$.
Note again that the low momentum behavior of (\ref{f_minimal}) may not be so simple because of the nontrivial time scale of the inflaton dynamics
during the transition from inflation to the reheating era.
As mentioned above, there are another contribution as (\ref{f_min_adi}), (\ref{f_conf_osc}) and (\ref{f_conf_adi}).
Again we stress that, in realistic situation, these contributions should be smoothly connected 
because we cannot strictly define the ``end'' of inflation and 
the typical time scale of the inflaton motion changes from $H_{\rm inf}$ to $m_\phi$ gradually around the transition epoch.
The number density is dominated by that from the oscillation effect (\ref{f_minimal}) for $m_\chi \gg H_{\rm inf}$
and two contributions are comparable for $m_\chi \lesssim H_{\rm inf}$.
In any case, the number density is given by
\begin{align}
	n_\chi(\tau) \sim \mathcal C H_{\rm inf}^3 \left( \frac{a_{\rm end}}{a(\tau)} \right)^3,
	\label{nchi_adi}
\end{align}
for $H_{\rm inf} < m_\chi < m_\phi$ where we numerically find $\mathcal C\sim 10^{-2}$. 
The present DM energy density from gravitational production divided by the entropy density is then given by
\begin{align}
	\frac{\rho_\chi^{\rm (GP)}}{s} \sim \frac{\mathcal C}{4} \frac{m_\chi H_{\rm inf}T_{\rm R}}{M_P^2}
	\simeq 3\times 10^{-10}\,{\rm GeV}\,\mathcal C\left(\frac{m_\chi}{10^9\,{\rm GeV}}\right)
	\left(\frac{H_{\rm inf}}{10^9\,{\rm GeV}}\right)
	\left(\frac{T_{\rm R}}{10^{10}\,{\rm GeV}}\right),
	\label{rho_GP}
\end{align}
for $m_\chi < m_\phi$ and $T_{\rm R}$ denotes the reheating temperature after inflation
and we assumed that the inflaton coherent oscillation behaves as non-relativistic matter.

%%%%%%%%%%%%%%%%%%%%%%%%%%%%%%%%
\subsection{Numerical simulation in realistic inflation model}
%%%%%%%%%%%%%%%%%%%%%%%%%%%%%%%%

Let us now calculate the phase space density numerically.
As a concrete inflation model, let us consider the hilltop inflation or new inflation models~\cite{Linde:1981mu,Albrecht:1982wi,Kumekawa:1994gx,Izawa:1996dv,Asaka:1999jb,Senoguz:2004ky,Boubekeur:2005zm}. 
The inflaton potential is given by
\begin{align}
	V(\phi) = M^4 \left[ 1 - \left(\frac{\phi}{v_\phi}\right)^n \right]^2,   \label{Vhill}
\end{align}
where $n\geq 6$ is favored from the recent cosmological data~\cite{Ema:2017rkk}. In a numerical calculation we take $n=6$.
This model is of our interest because it in general predicts a large hierarchy between $m_\phi$ and $H_{\rm inf}$.
Actually we have $m_\phi =\sqrt{2}nM^2/v_\phi  \gg H_{\rm inf} \simeq M^2/(\sqrt 3 M_P)$ for $v_\phi\ll M_P$.
The field value at the end of inflation is defined by 
\begin{align}
	\phi_{\rm end} = v_\phi\left(\frac{v_\phi}{\sqrt{2} n M_P}\right)^{1/(n-1)},
\end{align}
where the slow-roll parameter $\epsilon\equiv (M_P \partial V/\partial\phi)^2/(\sqrt 2V)^2$ becomes equal to unity.
The dimensionless power spectrum of the large scale curvature perturbation is given by
\begin{align}
	\mathcal P_\zeta \simeq \frac{\left[ 2n\left((n-2)N_e\right)^{n-1} \right]^{2/(n-2)}}{12\pi^2} \frac{M^4}{(v_\phi^n M_P^{n-4})^{2/(n-2)}},
\end{align}
where $N_e\simeq 50\mathchar`-\mathchar`-60$ denotes the e-folding number at which the perturbation with the present horizon scale exits the horizon.
The Planck observation suggests $\mathcal P_\zeta \simeq 2.2\times 10^{-9}$~\cite{Ade:2015lrj}.

\begin{figure}[t]
\begin{center}
\begin{tabular}{cc}
\includegraphics[scale=1.3]{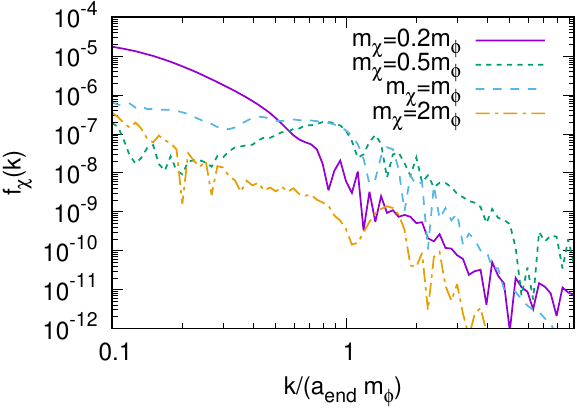}
\includegraphics[scale=1.3]{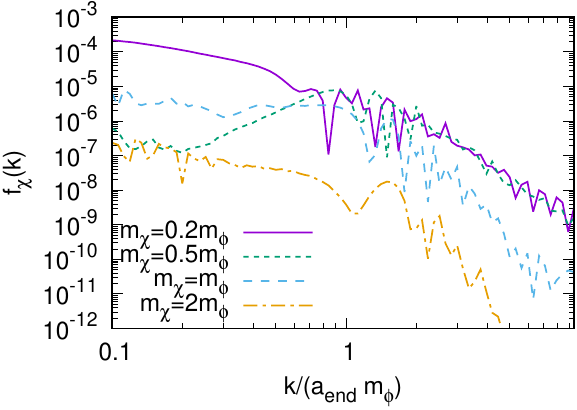}
\end{tabular}
\end{center}
\caption{
	The phase space density of $\chi$ after the gravitational particle production in hilltop inflation model with $v_\phi=0.5 M_P$.
	(Left) The case of conformal coupling $\xi=1/6$. We have taken $m_\chi = (0.2, 0.5, 1, 2) \times m_\phi$ for each line.
	(Right) The same as the left but for the minimal coupling $\xi=0$.
}
\label{fig:fchi}
\end{figure}

We numerically solved (\ref{alpha}) along with the background (\ref{phi_eom}) and (\ref{H2}).
Fig.~\ref{fig:fchi} shows the resulting phase space distribution $f_\chi(k)$ for the conformal coupling case $\xi=1/6$ (left)
and the minimal coupling case $\xi=0$ (right).
We have taken $v_\phi = 0.5 M_P$, which means $m_\phi /H_{\rm inf} \simeq 29$, and $M$ is taken to satisfy the observed density perturbation.
Three lines correspond to $m_\chi = (0.2,  0.5, 1, 2) \times m_\phi$, respectively.
The initial condition is set to be $\phi(\tau_i) \simeq 0.4 \phi_{\rm end}$ with $\tau_i = -1/(a(\tau_i) H_{\rm inf})$ 
where $a(\tau_i)$ can be arbitrary value.
The final evaluation time is taken to be $\tau_f = -\tau_i$.
The wave number in the horizontal axis is normalized by $a_{\rm end} m_\phi$ where $a_{\rm end}$ denotes the scale factor at
the end of slow-roll regime $\phi=\phi_{\rm end}$.
The results are consistent with (\ref{f_conf_osc}) for the conformal coupling case $\xi=1/6$ 
and (\ref{f_minimal}) for the minimal coupling $\xi=0$ for the high momentum mode $k \gtrsim a_{\rm end} m_\phi$.
In particular, the overall size of $f_\chi$ is almost the same as long as $m_\phi \gtrsim m_\chi$ 
for $\xi = 0$, and hence we conclude that the particles with $m_\phi \gtrsim m_\chi$ are indeed produced
even if $m_\chi \gg H_\mathrm{inf}$.
On the other hand, the overall size of $f_\chi$ decreases as $m_\chi$ increases beyond $m_\phi$
even for $\xi = 0$ (see the $m_\chi = 2m_\phi$ line in the right panel).
The lower momentum behavior comes mainly from the transition epoch from inflation to the oscillation epoch,
which is difficult to discuss analytically, but the estimations (\ref{rho_GP_conf}) and (\ref{rho_GP}) remain valid
because the number or energy density is dominated by the mode around $k\sim a_{\rm end} m_\phi$.\footnote{
	If $v_\phi$ is much smaller than $M_P$, the coherent oscillation just after inflation is not harmonic and the oscillation period
	can be much longer than $m_\phi^{-1}$. 
	Moreover, tachyonic instability makes the inflaton field inhomogeneous~\cite{Brax:2010ai}.
	These facts make robust calculation extremely difficult,
	although still the inflaton remains non-relativistic and we expect that the estimation here makes sense. 
}

%%%%%%%%%%%%%%%%
%%%%%%%%%%%%%%%%

%%%%%%%%%%%%%%%%
\begin{figure}[t]
\begin{center}
\begin{tabular}{cc}
\includegraphics[scale=1.3]{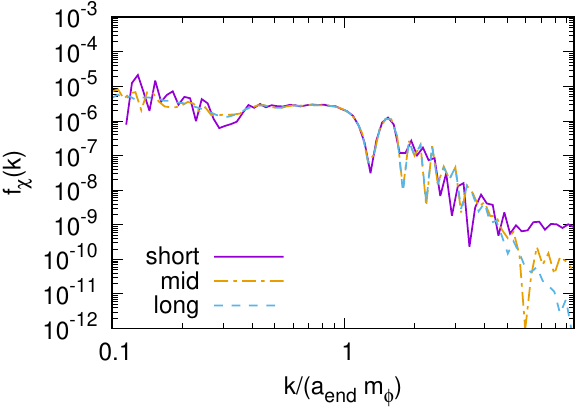}
\includegraphics[scale=1.3]{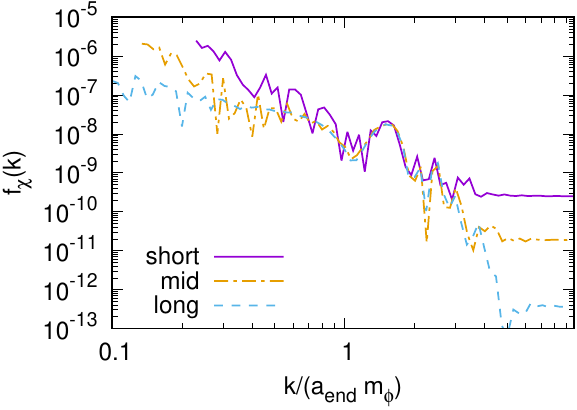}
\end{tabular}
\end{center}
\caption{
	The phase space density of $\chi$ after the gravitational particle production for the minimal coupling $\xi=0$.
	The three lines ``short'', ``mid'' and ``long'' correspond to the duration of inflation in a numerical calculation.
	We have taken $m_\chi = m_\phi$ (left) and $m_\chi = 2m_\phi$ (right).
}
\label{fig:f}
\end{figure}
%%%%%%%%%%%%%%%%

Now we comment on subtlety in evaluating $f_\chi(k)$.
As we discussed in the previous section, we formally define the particle number density
by taking $-\tau_i, \tau_f \rightarrow \infty$.
In our numerical calculation, however, we take some finite $\tau_i$ and  $\tau_f$
due to the limitation of computational cost,
and hence we should carefully subtract effects of finite $\tau_i$ and $\tau_f$.
More precisely, we want to take the limit $-k\tau_i, k\tau_f \to \infty$, since $\tau_i$ itself can be taken arbitrarily.
For fixed physical momentum $k / a_{\rm end}$, such a limit can be achieved by taking the 
duration of inflation long in a numerical calculation.
In order to identify such effects, we plot $f_\chi(k)$ for different initial condition in Fig.~\ref{fig:f}.
Three lines correspond to different initial condition, $\phi_i = (0.38, 0.41, 0.44) \times \phi_{\rm end}$ for ``long,'' ``mid'' and ``short,'' respectively.
The DM mass is taken to be $m_\chi = m_\phi$ (left) and $m_\chi = 2m_\phi$ (right).
The wave number in the horizontal axis is normalized by $a_{\rm end} m_\phi$.
As seen from the left panel, the flat part at large $k$ is initial condition dependent 
and becomes smaller as the initial time is taken to be earlier.
The following simple example may be helpful to understand this behavior.
Let us consider the integral:
\begin{align}
	I(k) = \int_{\tau_i}^{\tau_f} \frac{e^{ik\tau}}{\tau^2 + \tau_0^2}d\tau,
\end{align}
where $\tau_0$ is a real number.
In the limit $-\tau_i =\tau_f = \infty$, we can exactly solve it by using the residue theorem and obtain exponential form: $I(k) =\pi e^{-k\tau_0} / \tau_0$.
If we take large but finite $-\tau_i = \tau_f (\gg \tau_0)$, we instead have power law tail as
$I(k) \sim \pi e^{-k\tau_0} / \tau_0 + 1/(\tau_f^2 k)$.
The integrand is more complicated in a realistic situation,
but we expect that a similar phenomena occur in our numerical calculation.
Therefore the flat part at large $k$ is interpreted as an effect of finite $\tau_i$ and $\tau_f$, 
and we expect that it disappears for $-\tau_i, \tau_f \to -\infty$.
On the other hand, the modes with smaller $k$
are not affected by the change of the initial condition, 
and hence are expected to
be intact in the limit $-\tau_i, \tau_f \rightarrow \infty$.
We have checked that the result in Fig.~\ref{fig:fchi} 
is not affected by the change of the initial condition,
and hence we expect that it provides good estimation of $f_\chi$ 
in the limit $-\tau_i, \tau_f \rightarrow \infty$.
This issue is related to the choice of the initial condition at $\tau = \tau_i$.
If we could carefully choose the initial conditions of $\alpha_k$ and $\beta_k$ at $\tau=\tau_i$ so that they match the solution with $\alpha_k = 1$ and $\beta_k = 0$ at $\tau = -\infty$,
the $\tau_i$ dependence would be gone.

It is worth noting here that we do not actually know the initial condition of the Universe.
Inflation should last at least for $\sim 60$ e-foldings, but the dynamics before the ``observable'' inflation is unclear.
Therefore, strictly speaking, taking the limit $\tau_i\to -\infty$ may not be justified in the real Universe.
Similarly, the present Universe has of course finite age, and we cannot formally take $\tau_f\to \infty$.
We expect that taking these limits will give good estimation practically, but
it would be interesting that information of the initial condition of the Universe may be contained in the momentum distribution.

%%%%%%%%%%%%%%%%%%%%%%%%%%%%%%%%%%%%%%%%%%%%
\section{Discussion}  \label{sec:conc}
%%%%%%%%%%%%%%%%%%%%%%%%%%%%%%%%%%%%%%%%%%%%

We studied the gravitational production of scalar PGDM by tracing the evolution of its wave function from inflation to the reheating era.
We confirmed that PGDM mass with $m_\chi \gg H_{\rm inf}$ can be efficiently produced as long as $m_\chi \lesssim m_\phi$,
where $H_{\rm inf}$ and $m_\phi$ denote the Hubble scale during inflation and the inflaton mass, respectively.
Roughly, PGDM number density of $\sim H_{\rm inf}^3$ is produced even for $H_{\rm inf} \ll m_\chi$ $(\lesssim m_\phi)$
at the end of inflation and the beginning of the inflaton oscillation.
However, we note that our results depend on the assumption that the inflaton field remains homogeneous over the horizon scale
until the inflaton coherent oscillation becomes purely harmonic.
In actual situation, there can be some instability that tend to make the inflaton field inhomogeneous.
Although our results may not be affected much unless the inflaton field becomes highly relativistic, 
further investigation is needed in order to correctly estimate the PGDM abundance in various inflation models.

So far we have focused on the case of $m_\chi \gtrsim H_{\rm inf}$ in order to avoid the 
growth of large scale fluctuation during inflation.
If $m_\chi \ll H_{\rm inf}$, the superhorizon modes of the $\chi$ field are generated during inflation,\footnote{
	The condition of the growth of long wavelength mode in the dS era is $\nu^2 > 1/4$ in (\ref{chik_exact}), or
	$m_\chi^2 < (2-12\xi) H_{\rm inf}^2$.
}
and it will be effectively regarded as the homogeneous mode, which results in the coherent oscillation of $\chi$ field after inflation.
This is another source of the DM production.
Typical field variance averaged over the superhorizon mode is derived from the solution (\ref{chik_exact}) as~\cite{Linde:2005ht,Linde:2005yw}
\begin{align}
	\left< \chi^2\right> \simeq \frac{3 H_{\rm inf}^4}{8\pi^2 m_\chi^2}.  \label{chi2}
\end{align}
In a horizon patch, the $\chi$ field is almost homogeneous with its typical value given by $\sqrt{\left<\chi^2\right>}$. 
The $\chi$ field then begins a coherent oscillation when $H \sim m_\chi$.
The abundance of the coherent oscillation is estimated as
\begin{align}\label{eq:coherent}
	\frac{\rho_\chi^{\rm (CO)}}{s} \simeq \frac{T_{\rm R}}{8} \frac{\left< \chi^2\right>}{M_P^2}
	\simeq 8\times 10^{-12}\,{\rm GeV}\left( \frac{H_{\rm inf}}{10^9\,{\rm GeV}} \right)^4
	\left( \frac{10^9\,{\rm GeV}}{m_\chi}\right)^2
	\left( \frac{T_{\rm R}}{10^{10}\,{\rm GeV}} \right).
\end{align}
Here we assumed that $\chi$ begins to oscillate before the reheating is completed: $\Gamma_{\rm inf} < m_\chi$.
Otherwise, the abundance is suppressed by the factor $\sim \sqrt{m_\chi / \Gamma_{\rm inf}}$.
This also contributes to the DM density.
One crucial difference from the gravitationally produced contribution $\rho_\chi^{\rm (GP)}$ is that the large scale fluctuation of $\chi$ 
has an (uncorrelated) isocurvature perturbation, which is severely constrained from observation.\footnote{
	Note on terminology: coherent oscillation may also be regarded as gravitational production, since (\ref{chi2}) is induced
	gravitationally. We just call such long wavelength (superhorizon modes) contribution as coherent oscillation and 
	short wavelength (subhorizon modes) one as gravitational production.
}
The magnitude of DM isocurvature perturbation is estimated as
\begin{align}\label{eq:isoc}
	S_{\rm DM} \simeq R_\chi \frac{H_{\rm inf}}{\pi \sqrt{\left< \chi^2\right>}} = R_\chi \sqrt{\frac{8}{3}} \frac{m_\chi}{H_{\rm inf}},
\end{align}
where $R_\chi \equiv \rho_\chi^{\rm (CO)} / \rho_{\rm DM}$ denotes the fraction of 
$\chi$ coherent oscillation energy density in the total DM energy density.
The observational constraint is $S_{\rm DM} \lesssim 9\times 10^{-6}$~\cite{Ade:2015lrj}.
Note however that (\ref{chi2}) is an asymptotic averaged value when the inflation lasts long enough
and it may be possible to have larger/smaller field value in the actual Universe if $m_\chi \ll H_{\rm inf}$.
It is also affected by other terms like $\mathcal L \sim -\lambda \chi^4$.

Scattering of standard model particles in thermal bath also produces PGDM through the graviton exchange~\cite{Garny:2015sjg,Tang:2016vch,Tang:2017hvq,Garny:2017kha}.
The cross section for the process like $\overline\psi\psi \to \chi\chi$, where $\psi$ collectively denotes the standard model fields 
that are in thermal bath with temperature $T$, is
\begin{align}
	\sigma (\overline\psi\psi \to \chi\chi) \simeq  \mathcal{A}\frac{T^2}{M_P^4},
\end{align}
for $T \gg m_\chi$, $\mathcal{A}\simeq 1/24\pi, 1/48\pi $ and $1/12\pi$ for complex scalar, Dirac fermion and massless vector, respectively. The final DM abundance from thermal production is dominated by those produced in the earliest epoch and estimated as
\begin{align}
	\frac{\rho_\chi^{\rm (TP)}}{s} \sim \mathcal{A}\frac{m_\chi T_{\rm R} T_{\rm max}^8}{M_P^6 H_{\rm inf}^3}
	\sim  \mathcal{A} \frac{m_\chi T_{\rm R}^5}{M_P^4 H_{\rm inf}}
\end{align}
where $T_{\rm max} \sim (T_{\rm R}^2 H_{\rm inf} M_P)^{1/4}$ is the maximum cosmic temperature of the dilute plasma after inflation.
Compared with the inflaton-induced gravitational contribution (\ref{rho_GP}), 
the thermal production contribution is suppressed by $T_{\rm R}^4 / (H_{\rm inf}^2 M_P^2)$.
Thus thermal contribution is subdominant in most cases, and some DM scenarios implementing thermal production solely were discussed in Ref.~\cite{Aoki:2016zgp, Babichev:2016bxi, Kolb:2017jvz, Chen:2017kvz, Bernal:2018qlk}, for examples. However, it is possible that $\chi$ is much heavier than the inflaton, $m_\chi \gg m_\phi$, so that all the gravitational production is not effective, while $T_{\rm max} > m_\chi$ and hence thermal production is active. In such a case, thermal production can give a dominant contribution to the PGDM abundance.

%%%%%%%%%%%%%%%%
\begin{figure}[t]
	\begin{center}
		\begin{tabular}{cc}
			\includegraphics[scale=0.78]{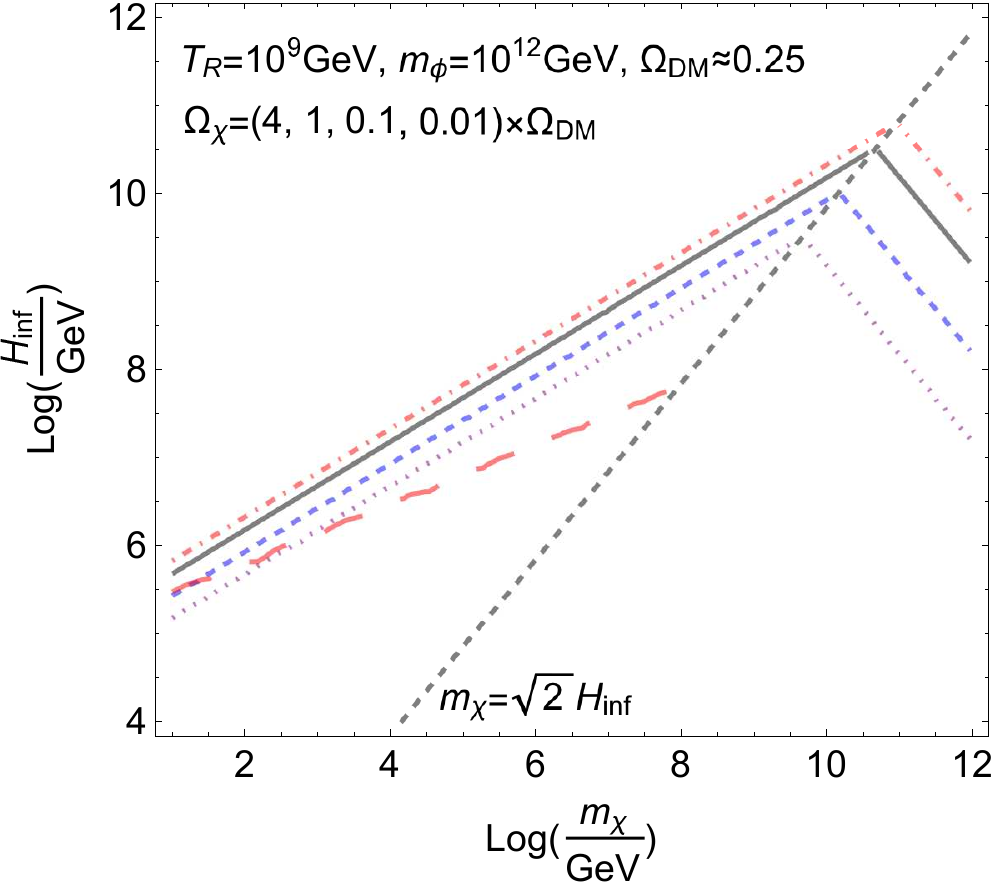}
			\includegraphics[scale=0.78]{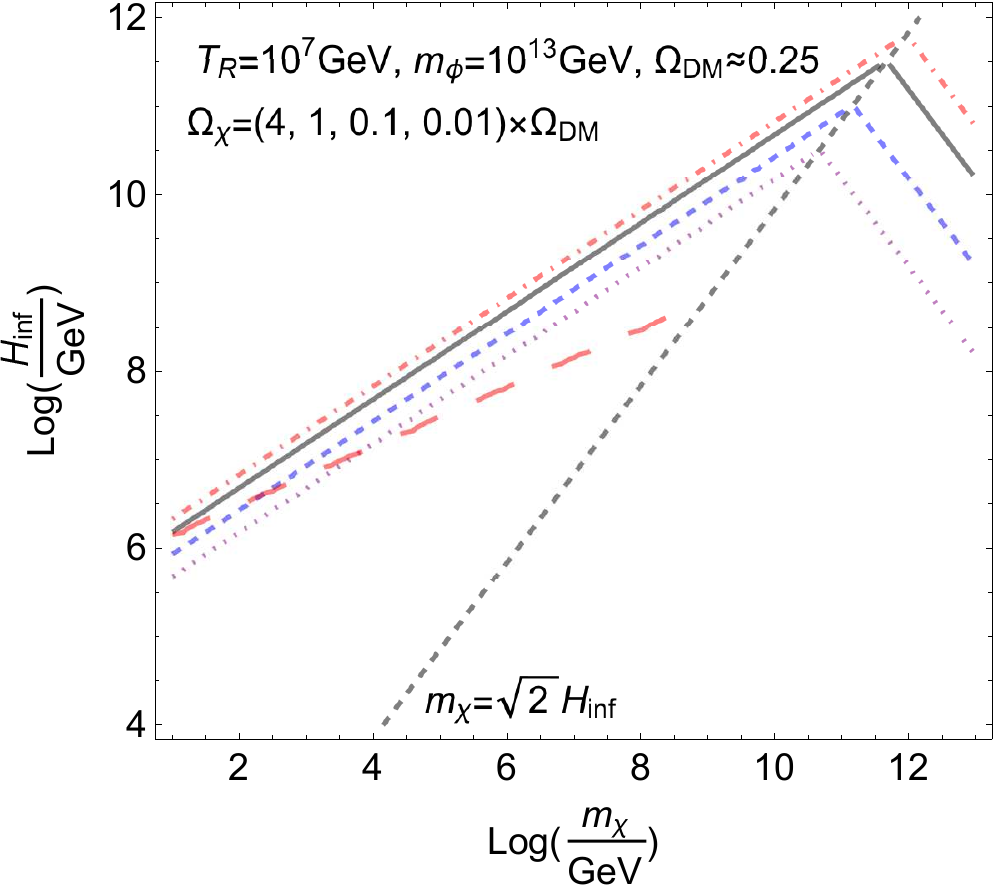}
		\end{tabular}
	\end{center}
	\caption{Illustration with reheating temperature $T_{\rm R}=10^{9}\,$GeV $(10^7\,{\rm GeV})$ 
	and inflaton mass $m_\phi=10^{12}\,$GeV $(10^{13}\,{\rm GeV})$ for minimal coupling $\xi=0$ in the left (right) panel. 
	Four different lines (dot-dashed, solid, dashed, and dotted) correspond to $\Omega_{\chi}=(4,1,0.1,0.01)\times\Omega_{\textrm{DM}}$. The long dashed red line shows the isocurvature perturbation limit and region above this line is excluded. 
	The dominant production mechanism is coherent oscillation for $m_{\chi} < \sqrt 2 H_{\rm inf}$ and gravitational production for $m_{\chi} > \sqrt 2 H_{\rm inf}$.
	\label{fig:sum}	}
\end{figure}
%%%%%%%%%%%%%%%%

In Fig.~\ref{fig:sum}, we show the total contribution for PGDM's relic abundance $\Omega_{\chi}$ as functions of $H_{\textrm{inf}}$ and DM mass $m_{\chi}$. We have chosen the reheating temperature $T_{\rm R}=10^{9}\,$GeV $(10^7\,{\rm GeV})$ 
and inflaton mass $m_\phi=10^{12}\,$GeV $(10^{13}\,{\rm GeV})$ for minimal coupling $\xi=0$ in the left (right) panel. 
From top to down, different lines (dot-dashed, solid, dashed, and dotted) correspond to $\Omega_{\chi}=(4,1,0.1,0.01)\times\Omega_{\textrm{DM}}$. The long dashed line marks the isocurvature perturbation limit (\ref{eq:isoc}) and region above this line is excluded. 
Note that the dominant production mechanism is coherent oscillation for $m_{\chi} < \sqrt 2 H_{\rm inf}$
and gravitational production for $m_{\chi} > \sqrt 2 H_{\rm inf}$.
In the shown parameter space, the thermal production can always be neglected. 
DM in the shown region might be probed by cosmic ray and CMB experiments if there is (small) $Z_2$ breaking term and it decays, see Ref.~\cite{Tang:2016vch,Ibarra:2013cra}.

Finally we briefly comment on the case of large nonminimal coupling $\xi$.
Compared with the case of minimal coupling $\xi=0$, the PGDM abundance is expected to scale as $(\xi-1/6)^2$.
However, for very large $\xi \gtrsim m_\phi^2 / H_{\rm inf}^2$,
the $\chi$ field feels large tachyonic mass because the Ricci scalar is oscillating between positive and negative value during inflaton oscillation
and low momentum modes are exponentially enhanced due to the tachyonic instability~\cite{Bassett:1997az,Tsujikawa:1999jh,Markkanen:2015xuw}.
In such a case the final PGDM abundance may be greatly enhanced.

%%%%%%%%%%%%%%%%%%%%%%%%%%%%%%%%%%%%%%%%%%%%
\section*{Acknowledgments}
%%%%%%%%%%%%%%%%%%%%%%%%%%%%%%%%%%%%%%%%%%%%

The work of Y.E. was supported in part by JSPS Research Fellowships for Young Scientists.
This work was supported by the Grant-in-Aid for Scientific Research C (No.18K03609 [KN]), Scientific Research A (No.26247042 [KN]), Young Scientists B (No.26800121 [KN]) and Innovative Areas (No.26104009 [KN], No.15H05888 [KN], No.17H06359 [KN], No.16H06490 [YT]).

\appendix

%%%%%%%%%%%%%%%%%%%%%%%%%%%%%%%%%%%%%%%%%%%%%%
\section{Gravitational particle production in a toy expansion model with conformal coupling}
\label{sec:toy}
%%%%%%%%%%%%%%%%%%%%%%%%%%%%%%%%%%%%%%%%%%%%%%

Here we calculate the gravitational particle production caused purely by the Hubble expansion of the Universe,
i.e., without the effect of inflaton oscillation, to check the estimation of Sec.~\ref{sec:conf}.
In order to get rid of the effect of inflaton oscillation, 
let us consider the toy model in which the $\phi$ energy density goes as 
\begin{align}
	\rho_\phi(\tau) = \frac{\rho_{\phi,i}}{ 1 + (a(\tau) / a_{\rm end})^3}.
	\label{rhophi_toy}
\end{align}
Inflation happens for $a(\tau) \ll a_{\rm end}$ and it smoothly connects to the MD Universe for $a(\tau) \gg a_{\rm end}$
and $\rho_{\phi,i} = 3H_{\rm inf}^2 M_P^2$.

Let us consider the integral (\ref{betak}) in this simplified model with the conformal coupling $\xi=1/6$.
A general rule is that, since $\omega_k' / \omega_k$ is a smooth function with its typical time scale at most $\sim \mathcal H^{-1} \sim \tau$,
if the time scale of the phase factor in the integrand is much faster, the integral is exponentially suppressed as $\sim e^{-k \tau}$.
Actually $\omega_k' / \omega_k$ changes its slope at $t\simeq t_k$ and $t\simeq t_{\rm end}$
and the whole integral is dominated around these epochs. 
First let us consider the low momentum mode $k < a_{\rm end} m_\chi$.
In this case we have $t_k < t_{\rm end}$ and contributions around $t\simeq t_k$ and $t\simeq t_{\rm end}$ are comparable.
As a result, $f_\chi(k) = |\beta_k|^2$ is given as
\begin{align}
	f_\chi(k) \sim \exp\left( -\frac{cm_\chi}{H_{\rm inf}} \right),  \label{fk_toy_lowk}
\end{align}
where we have introduced a numerical factor $c$. We will see that $c\sim 5$ fits the numerical results well.
Next let us consider the high momentum mode $k > a_{\rm end} m_\chi$.
In this case we have $t_k > t_{\rm end}$ and we obtain
\begin{align}
	f_\chi(k) \sim \frac{a_{\rm end}^4 m_\chi^4}{k^4}\exp\left( -\frac{ck}{a_{\rm end}H_{\rm inf}} \right)
	+ \exp\left[ -c\left(\frac{k}{k_c} \right)^{3/2} \right],
	\label{fk_toy_highk}
\end{align}
where the first (second) term corresponds to the integral around $t\sim t_{\rm end}$ $(t\sim t_k)$,
and $k_c \equiv a_{\rm end} H_{\rm inf} (m_\chi / H_{\rm inf})^{1/3}$ is defined by the momentum so that $m_\chi t_k = 1$.
If $m_\chi < H_{\rm inf}$, the second term mostly dominates.
We have omitted the mixing term since it is subdominant in most cases.
Intuitively, modes with $k> k_c$ are always adiabatic against the expansion of the Universe.
If $m_\chi > H_{\rm inf}$, the first term dominates
and all the modes are suppressed by at least the factor $\sim e^{-m_\chi / H_{\rm inf}}$,
since they are all adiabatic throughout the whole history of the Universe and no significant excitations are expected.
As a result, the number density due to the gravitational production in this case is approximated as
\begin{align}
	n_\chi(\tau) \sim \mathcal A\left( \frac{a_{\rm end}}{a(\tau)} \right)^3 \times \begin{cases}
		m_\chi H_{\rm inf}^2 & {\rm for}~~~m_\chi \lesssim H_{\rm inf} \\
		e^{-cm_\chi/ H_{\rm inf}} m_\chi^3   & {\rm for}~~~m_\chi \gtrsim H_{\rm inf}
	\end{cases},
\end{align}
where we find that the numerical factor $\mathcal A$ is weakly dependent on $m_\chi$ and 
$\mathcal A \sim 6\times 10^{-4}$ for $m_\chi \lesssim H_{\rm inf}$.

%%%%%%%%%%%%%%%%
\begin{figure}[t]
\begin{center}
\begin{tabular}{cc}
\includegraphics[scale=1.5]{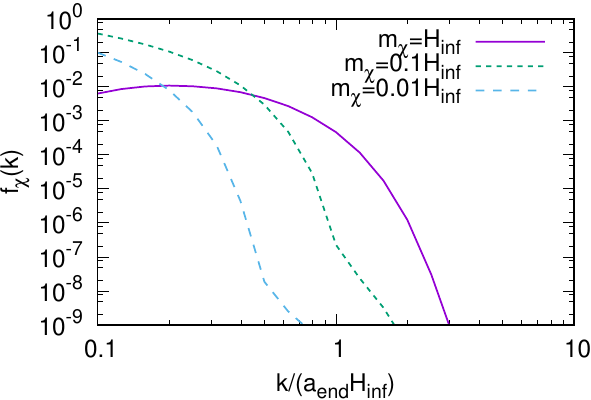}
\end{tabular}
\end{center}
\caption{
	The phase space distribution of $\chi$ after the gravitational particle production
	in a toy model (\ref{rhophi_toy}) with conformal coupling $\xi=1/6$,  
	for $m_\chi = (1,\,0.1,\,0.01)\times H_{\rm inf}$.
}
\label{fig:toy}
\end{figure}
%%%%%%%%%%%%%%%%

We solved (\ref{alpha}) with the Friedmann equation $H^2 = \rho_\phi/(3M_P^2)$
by using the toy model (\ref{rhophi_toy}) for the conformal coupling $\xi=1/6$ to estimate the gravitational particle production rate.
The result of $f_\chi(k)$ is shown in Fig.~\ref{fig:toy} for $m_\chi = (1,\,0.1,\,0.01)\times H_{\rm inf}$.
The results are well fitted by the formula (\ref{fk_toy_lowk}) and (\ref{fk_toy_highk}) with $c=5$.

%%%%%%%%%%%%%%%%%%%%%%%%%%%%%%%%%%%%%%%%%%%%%%%%%%

%%%%%%%%%%%%%%%%%%%%%%%%%%%%%%%%%%%%%%%%%%%%%%%%%%

\end{document}